\documentclass[preprint,authoryear,12pt]{elsarticle}

\usepackage{geometry}
\usepackage{graphicx} 
\usepackage{amsmath, amssymb, amsfonts, amsthm, float} 
\usepackage{enumerate, color, framed, float, multirow}
\usepackage{comment, longtable, caption, subcaption, appendix}
\usepackage{setspace, parskip}
\usepackage{placeins}

\biboptions{longnamesfirst, sort, authoryear}
\graphicspath{{Graphics/}}

\allowdisplaybreaks

\newcommand{\R}{\mathbb{R}}

\newtheoremstyle{quest}{\topsep}{\topsep}{}{}{\bfseries}{}{ }{\thmname{#1}\thmnote{ #3}.}

\journal{Social Networks Journal}

\begin{document}
\title{Ensuring Reliable Monte Carlo Estimates of Network Properties}
 \author[1]{Haema Nilakanta\corref{cor1}}  
 \ead{nilak008@umn.edu}
\author[2]{Zack W. Almquist} \ead{zalmquis@uw.edu}
\author[1]{Galin L. Jones \fnref{fn1}} \ead{galin@umn.edu}

\cortext[cor1]{Corresponding author}
 \fntext[fn1]{Partially supported by NSF grant \#1922512}
\address[1]{School of Statistics, University of Minnesota, Twin Cities, 313 Ford Hall
224 Church St SE, Minneapolis, MN 55455}
\address[2]{Department of Sociology and eScience, University of Washington}

\begin{abstract}
The literature in social network analysis has largely focused on methods and models
  which require complete network data; however there exist many
  networks which can only be studied via sampling methods due to the scale or complexity of the network, access limitations, or
  the population of interest is hard to reach. In such cases, the application of random walk-based Markov chain Monte Carlo (MCMC) methods to estimate multiple network features is common. However, the reliability of these estimates has been largely ignored. We consider and further develop multivariate MCMC output analysis methods in the context of network sampling to directly address the reliability of the multivariate estimation. This approach yields principled, computationally efficient, and broadly applicable methods for assessing the Monte Carlo estimation procedure. In particular, with respect to two random-walk algorithms, a simple random walk and a Metropolis-Hastings random walk, we construct and compare network parameter estimates, effective sample sizes, coverage probabilities, and stopping rules, all of which speaks to the estimation reliability.
\end{abstract}



\maketitle


\section{Introduction and Background}
\label{sec:intro}

Much of the network literature has focused on complete network data \citep{scott2017social, wasserman1994social, Kolaczyk2009}; but in many practically relevant settings, the full network is difficult to study due to its scale or complexity (e.g., geospatial social networks) or the network represents a hidden population (e.g., homeless friendship networks in United States). In such cases, traditional survey sampling methods, e.g., simple random sampling (SRS), are not practical due to the absence of a sampling frame. Alternatively, one can collect an approximately uniform sample from a network by traversing the structure in a nondeterministic manner. Features of interest can then be estimated using sample statistics. A particular focus within the network sampling literature is on traversing networks with random walk-based algorithms, a Markov chain Monte Carlo (MCMC) method.

 Overall, there are three core approaches to sample networks: (1) SRS of nodes, also known as egocentric sampling \citep{wasserman1994social}, (2) sampling  edges at random, or (3) MCMC sampling, commonly referred to as traversal sampling or link trace sampling. Practical advantages to each of these methods exist, although SRS and link trace sampling are more common than edge sampling. There exists an extensive literature which looks at SRS and other more complex sampling designs (e.g., cluster or stratified designs) in the social network literature \citep{marsden2011survey}.

 Link trace sampling methods are particularly popular in the social sciences. With these traversal approaches, one can leverage the underlying network structure without a sampling frame to obtain population level measures. One of the most popular versions is Respondent Driven Sampling (RDS); first introduced in \cite{heckathorn1997respondent}. Since its introduction, there have been several extensions of RDS that further underline the appeal of link trace methods to study social interactions \citep[see e.g.,][]{salganik2004sampling,gile2010respondent,handcock2010modeling}. More recently, the growth of large Online Social Networks (OSNs) has also brought rising attention to traversal methods. For example, \cite{gjoka2010walking} and others \citep[e.g.,][]{kurant2012coarse,gjoka2011multigraph} have used these methods to obtain asymptotically unbiased estimates of core network features (e.g., the degree distribution or clique census) or subgroup populations. 

Among large OSNs, random walk-based algorithms have been regularly used to estimate key features such as average connectedness or clustering coefficients \citep{gjoka2011practical}. These random walk algorithms have also been employed to obtain information about hard to reach populations, such as estimating disease prevalence among individuals at high risk for HIV \citep{thompson:2017}. While the use of these MCMC methods to estimate network features is common, the quality of estimation with these Monte Carlo samples has not been directly addressed in a computationally efficient way. We contribute to this area by considering and further developing MCMC output analysis methods in the context of network sampling that directly address the reliability of estimation.

Constructing MCMC sampling algorithms to efficiently traverse a network can be challenging and is an active area of research. As a result, there has been substantial work on comparing various MCMC
sampling methods for networks, but the comparisons usually only consider
the properties of univariate point estimates, computation speed (i.e.,
clock time or percent of network sampled), or the difference in empirical distributions using the
Kullback-Leibler divergence, Kolmogorov-Smirnov D-statistic, or the
total variation distance \citep[see, among
others,][]{lee2006statistical, avrachenkov2018revisiting,
  gjoka2011practical, gile2010respondent, blagus2017empirical,
  ahmed2014network, zhou2016faster, li2015random,
  leskovec2006sampling, wang2011understanding,
  salamanos2017deterministic, lee2012beyond,
  joyce2011kullback}. Typically the goal is to
estimate many network features based on one Monte Carlo sample, while
comparisons typically focus on univariate summaries. That is, the
multivariate nature of the estimation problem has been broadly
ignored.

Moreover, separate from the natural variability in the data, the estimates produced by these Monte Carlo methods are also subject to Monte Carlo error in that different runs of the sampling algorithm will result in different estimates. Thus, the algorithm used will impact the quality of the estimation. Of course, if the Monte Carlo sample sizes are large enough, then the differences in run estimates will be negligible.  This then raises the question, how large is large
enough?  That is, how large does the Monte Carlo sample need to be so
that the estimates are trustworthy? 

The current tools used in the network sampling literature to determine when to terminate the sampling process are insufficient. Popular methods rely on the use of so-called convergence diagnostics \citep{cowl:carl:1996,
  gelm:rubi:1992, gewe:1992, heid:welc:1983}, but none of these
methods make any attempt to assess the quality of estimation \citep{fleg:etal:2008, jone:etal:2006}. Moreover, these diagnostics have been shown to stop the sampling process prematurely \citep{jone:etal:2006,vats:2018revisiting}. Another common approach is to study the running mean plot and determine the point at which it stabilizes to find approximately when the estimates have settled \citep{gjoka2011practical,lu2012sampling,lee2006statistical,ribeiro:2010}. This approach is inadequate since its interpretation is subject to how much one zooms in on a section of the plot.

Although the network sampling literature on Monte Carlo estimation reliability is relatively sparse, \cite{Avrachenkov2016,lee2006statistical,chiericetti2016sampling,salamanos2017deterministic}, and \cite{wang2011understanding} considered the relative error or normalized root mean squared error of sample estimates from various sampling methods. However, neither approach takes into account the multivariate nature of the problem nor tries to calculate the sample variance from the correlated sampling procedure. In addition, \cite{mohaisen:2010} and \cite{zhou2016faster} discuss the theoretical mixing time of the sampling algorithms they propose, although theoretically valid, are impractical to implement. We are unaware of any other work that directly address the reliability of the multivariate estimation with these MCMC samples.

We consider and further develop multivariate MCMC output
analysis methods \citep[see e.g.][]{vats2015multivariate,
  vats2018strong, flegal2015mcmcse} in the context of network
sampling with respect to two MCMC algorithms: a simple random walk and a random walk-based version of the Metropolis-Hastings
algorithm. This approach yields principled, computationally efficient,
and broadly applicable methods for assessing the reliability of the
Monte Carlo estimation procedure.  In particular, we construct and
compare network parameter estimates, effective sample sizes, coverage
probabilities, and stopping rules.

The rest of the paper is organized as follows. In Section 2 we introduce some basic network notation and MCMC methods on networks. We also introduce output analysis tools to determine multivariate MCMC estimation reliability. In Sections 3 and 4 we further develop these output analysis tools in the context of network sampling, providing three examples of their use on a simple simulated high school social network to illustrate the concepts and progressively move to more complicated, larger networks. Finally, we conclude with a discussion in Section 5.


\section{Methods}
\subsection{Markov Chain Monte Carlo Methods on Networks}\label{Monte Carlo Methods on Networks}

We represent the network of interest in terms of a graph \citep[see][]{wasserman1994social}, which is a relational
structure comprised of two elements: a set of nodes or vertices (used
interchangeably), and a set of vertex pairs representing edges or ties (i.e., a
relationship between two nodes). Formally, let $V$ denote a non-empty countable set of nodes,
$E \subseteq V \times V$ denote the set of edges between the
vertices, and $G=(V,E)$ denote the network. We only
consider simple networks that are binary, undirected, well-connected, and without self
loops.  Define the network size, $n$, to be the set cardinality of
$V$. Similarly, $n_e$ is the number of edges in the graph.  The
network features of interest can be expressed as the mean
of a function over the entire network. More formally, suppose
$h: V \rightarrow \R^p$ where $p$ is the number of features of
interest and let $\lambda$ be the uniform distribution on $V$.  Then,
if $X \sim \lambda$, we want to calculate the $p$-dimensional mean
vector
\begin{equation}
\label{eq:network means}
 E_{\lambda}[h(X)] = \frac{1}{n} \sum\limits_{ v \in V} h(v),
\end{equation}
where the subscript indicates that the expectation is calculated with
respect to $\lambda$.  It will be notationally convenient to denote
$E_{\lambda}[h(X)] = \mu_h$ and we will use both
interchangeably. Specific network features of interest might include:
mean degree, degree distribution, mean clustering coefficient, and
proportion of nodes with specific nodal attributes, e.g., proportion of female users in an OSN.

Computing $\mu_h$ is often difficult in practically relevant
applications and hence we turn to MCMC methods.  Let
$\{V_0, V_1, V_2, \ldots \}$ be an irreducible, aperiodic Markov chain
with invariant distribution $\lambda$ \citep[for definitions
see][]{bremaud_2010, levin2009markov}.  Then by Birkhoff's ergodic
theorem we have that, if $E_{\lambda}|h(X)| < \infty$, with
probability 1,
\begin{equation}
\label{eq:slln}
\mu_m = \frac{1}{m} \sum_{t=0}^{m-1} h(V_t) \to \mu_h, ~\text{as } ~ m \to \infty. 
\end{equation}
Thus estimation of $\mu_h$ is straightforward; simulate $m$ steps of
the Markov chain and use the sample mean.  However, the quality of estimation depends on the Monte Carlo 
sample size, $m$, since for a finite $m$ there will be an unknown
\textit{Monte Carlo error}, $\mu_m - \mu_h$.  We can begin to assess
this error through a central limit theorem \citep[see
e.g.][]{aldous1997mixing, jones2004markov, vats2015multivariate}.
That is, for any initial distribution of the Markov chain, as
$m \to \infty$,
\begin{equation}
\label{eq:clt}
\sqrt{m}(\mu_m - \mu_h) \stackrel{d}{\to} N_p(0, \Sigma),
\end{equation}
where
\begin{equation}
\label{eq:sigma}
\Sigma = \text{Var}_\lambda(h(V_0)) + \sum\limits_{t=1}^\infty \left[\text{Cov}_\lambda (h(V_0), h(V_t)) + \text{Cov}_\lambda (h(V_0), h(V_t))^T \right].
\end{equation}
If $\|\cdot\|$ denotes the standard Euclidean norm, then, given our
assumptions on the Markov chain, the main requirement for
\eqref{eq:clt} is that $E_{\lambda}[\|h\|^2] < \infty$, which typically will hold. Also, since the chain is on the finite state space $V$, it is uniformly ergodic. \citep{aldous1997mixing}. 

The matrices $\Sigma$ and $\Lambda := \text{Var}_\lambda(h(V_0))$ will
be fundamental to the remainder. Estimating $\Lambda$ is
straightforward using the sample covariance, denoted $\Lambda_m$, but
estimating $\Sigma$ is a nontrivial matter which has attracted a
significant research interest \citep{andr:1991, chen1987multivariate,
  dai2017multivariate, liu:fleg:2018, liu:fleg:2018spec,
  hobe:etal:2002, jones2006fixed, kosorok2000monte, seil:1982,
  vats2018strong, vats2015multivariate, vats:fleg:2018}. There are several approaches to estimate $\Sigma$ that use spectral variance estimators, but these are computationally demanding especially with large Monte Carlo sample sizes \citep{liu:fleg:2018} . Therefore due to computational feasibility, 
  we will only consider the method of batch means, which we present now. 

Let
$\{X_t, \, t \ge 0\} = \{h(V_t), \, t \ge 0\}$ and set $m = a_m b_m$
where $a_m$ is the number of batches and $b_m$ is the batch size. For
$k = 0, \ldots, a_m-1$ set

\[
  \bar{X}_k := b_m^{-1} \sum_{t=0}^{b_m-1} X_{k b_m +t}.
\]
Then $\bar{X}_k$ is the mean vector for batch $k$ and the estimator of
$\Sigma$ is
\[
  \Sigma_m = \frac{b_m}{a_m-1} \sum\limits_{k=0}^{a_m-1} (\bar{X}_k -\mu_m)(\bar{X}_k - \mu_m)^T.
\]
For $\Sigma_m$ to be positive definite, $a_m > p$. It is common to choose $a_m = \lfloor m^{1/2} \rfloor$ or
$a_m = \lfloor m^{1/3} \rfloor$ where $a_m > p$ is met.  Batch means produces a strongly consistent
estimator of $\Sigma$ \citep{vats2015multivariate} under conditions
similar to those required for \eqref{eq:clt} and is implemented in the
\texttt{mcmcse} R package \citep{flegal2015mcmcse}.

\subsection{MCMC Output Analysis}
\label{sec:output}

It would be natural to use the CLT and $\Sigma_m$ to form
asymptotically valid confidence regions for $\mu_h$. The volume of the
confidence region could then be used to describe the precision in the
estimation and, indeed, this sort of procedure has been advocated
\citep{jone:etal:2006}. More specifically, if $T^2_{1-\alpha, p, q}$
denotes the $1-\alpha$ quantile of a Hotelling's $T$-squared
distribution where $q = a_m - p$, then a $100(1-\alpha)$\% confidence ellipsoid for
$\mu_h$ is the set
\[
  C_\alpha(m) = \{\mu_h \in \R^p: m(\mu_m -\mu_h)^T \Sigma_m^{-1}(\mu_m - \mu_h) < T^2_{1-\alpha, p, q}\} .
\]
The volume of the ellipsoid is given by
\[
  \text{Vol}(C_\alpha(m)) = \frac{2\pi^{p/2}}{p\Gamma(p/2)} \left(\frac{T_{1-\alpha, p, q}}{m}\right)^{p/2} |\Sigma_m|^{1/2}.
\]
One could then terminate a simulation when the volume is sufficiently
small, indicating that our Monte Carlo error is sufficiently low.  
However, the fixed-volume approach is difficult to
implement even when $p$ is small \citep{flegal2015mcmcse,vats2015multivariate,glynnwhitt:1992}.

An alternative is to terminate the simulation when the volume is small
compared to the generalized variance \citep{wilk:1932} of the target distribution, that is, if
$|\cdot|$ denotes determinant, small compared to $|\Lambda|$.  The
intuition is that when the Monte Carlo error is small compared to the
variation in the target distribution, then it is safe to stop.  More
formally, letting $m^* > 0$ and $\epsilon >0$ be given, then we
terminate the simulation at the random time $T_{SD}(\epsilon)$ defined as,
\[
  T_{SD}(\epsilon) = \inf\left\{m \ge 0: \text{Vol}(C_\alpha(m))^{1/p} + \epsilon |\Lambda_m|^{1/2p}I(m < m^*) + m^{-1} \le \epsilon |\Lambda_m|^{1/2p}\right\}.
\]
The role of $m^*$ is to require some minimum simulation effort. It should be large enough so that both $\Lambda_{m^*}$ and $\Sigma_{m^*}$
are positive definite and the lower bound on the ESS is achievable.

We can connect $T_{SD}(\epsilon)$ to effective sample size, the equivalent number of independent and identically distributed (\textit{iid}) samples that would give the same standard error as the correlated sample,
\begin{equation}
\label{eq:ESS}
\text{ESS}  = m \left[\frac{| \Lambda |}{ |\Sigma| }\right]^{1/p}
\end{equation}
and naturally estimated with
\begin{equation}
\widehat{\text{ESS}}  = m \left[\frac{| \Lambda_m |}{ |\Sigma_m| }\right]^{1/p}. \label{estimatedessequation}
\end{equation}
By rearranging the defining inequality of $T_{SD}(\epsilon)$ we see
that terminating at $T_{SD}(\epsilon)$ is essentially equivalent, for
large $m$, to terminating when the estimated effective sample size
satisfies
\[
 \widehat{\text{ESS}} \ge \frac{2^{2/p}\pi}{(p\Gamma(p/2))^{2/p}}
 \frac{\chi^2_{1-\alpha, p}}{\epsilon^2}. 
 \]
 Notice that the right-hand side of the inequality can be calculated
 prior to running the simulation and hence yielding a minimum
 simulation effort based on a desired confidence level $1-\alpha$ and
 relative precision $\epsilon$.

 Later, we will require the delta method \citep[see
 e.g.][Ch. 3]{sen:sing:1993}.  This substantially broadens the
 application of the methodology so far described.  We are often
 interested in estimating $g(\mu_h)$ where $g: \R^p \rightarrow \R^p$.
 If $g$ is such that it has a non-null derivative $\nabla g(\mu_h)$ at
 $\mu_h \in \R^p$ and is continuous in a neighborhood of $\mu_h$,
 then, as $m \to \infty$, the strong law at \eqref{eq:slln} ensures
 $g(\mu_m) \to g(\mu_h)$, with probability 1, and the CLT at
 \eqref{eq:clt} ensures that
 \begin{equation}
   \label{eq:dm.clt}
 \sqrt{m} (g(\mu_m) - g(\mu_h)) \stackrel{d}{\to} \text{N}\left(0,
   [\nabla g(\mu_h)]^T \Sigma [\nabla g(\mu_h)] \right) .
\end{equation}
It is straightforward to estimate the asymptotic covariance with
\[
 [\nabla g(\mu_m)]^T \Sigma_m [\nabla g(\mu_m)].
\]
Thus we can proceed with the output analysis as described above.
Notice that
\begin{equation}
  \label{prop:dESS}
  \text{ESS}_{g} :=  m \left[\frac{|[\nabla g(\mu_h)]^T \Lambda
      [\nabla g(\mu_h)]|}{|[\nabla g(\mu_h)]^T \Sigma [\nabla
      g(\mu_h)] |}\right]^{1/p} = m \left[\frac{| [ \nabla
      g(\mu_h)]^T| |\Lambda| |[\nabla g(\mu_h)]|}{|[\nabla
      g(\mu_h)]^T|   |\Sigma|| [\nabla 
      g(\mu_h)] |}\right]^{1/p} = \text{ESS}  
\end{equation}
and hence ESS is unaffected by the delta method transformation.

\subsection{Two MCMC Sampling Methods} \label{RW sampling methods}

We will consider two random walk-based MCMC methods, a
simple random walk (SRW) and a
Metropolis-Hastings (MH) algorithm with a simple random walk proposal. MH is constructed to have its invariant
distribution as $\lambda$, the uniform distribution over nodes.
SRW has a different invariant distribution, necessitating the
use of importance sampling in estimation.  The details are considered
below.

First, we require some notation. If there is an edge  from node $i$ to node $j$ we say $i$ and $j$ are
neighbors. The number of neighbors of node $i$ is its degree, $d_i$.

Then the SRW works as follows, if the current state is $i$, then the transition
probability of moving to node $j$ is 
\[
  P(i,j)^{SRW} = \begin{cases} 
  \frac{1}{d_i} &\quad \text{if $j$ is a neighbor of $i$}\\
  0 &\quad \text{otherwise.}
  \end{cases}
\]
The stationary density of the SRW is $\lambda^*(i) = d_i / 2 n_e$, which is not the uniform. 

\cite{gjoka2011practical} suggested using a Metropolis-Hastings
algorithm with SRW as the proposal distribution (for a summary of the 
Metropolis-Hastings algorithm refer to \citep{bremaud_2010}).  This gives rise to
MH transition probabilities of the form
\[
  P(i,j)^{MH} = \begin{cases}
  \frac{1}{d_i} \min\left(1, \frac{d_i}{d_j}\right) \quad &\text{if $j$ is a neighbor of $i$}\\
  1 - \sum_{k\ne i} \frac{1}{d_i} \min\left(1, \frac{d_i}{d_k}\right) \quad &\text{if $j =i$}\\
  0 \quad &\text{otherwise.}
  \end{cases}
\] 
In this case, the stationary density is the uniform over $V$,
$\lambda(i) = 1/n$.


\section{Monte Carlo Methods for Network Descriptive Statistics and Inference}\label{Simulations}

We focus on estimating popular network features, these include: mean degree, degree
distribution (e.g., proportion of nodes with $k$ neighbors), mean clustering coefficient, and mean of nodal
attributes.  For a given node $v$, let $d_v$ be the degree, $t_v$ be
the number of triangles, and a categorical attribute, $x_v$,
(e.g., race) having $c$ levels $x(1), x(2), \ldots, x(c)$. We keep these estimators general as one can easily see that the list can be expanded. In terms of
the notation from the previous section where $\mathbb{I}$ denotes the indicator function, we want to estimate $\mu_h$
where
\begin{equation}
  \label{eq:newh}
h(v) = (d_v,\, \mathbb{I}(d_v = k), \, 2t_v \mathbb{I}(d_v \ge 2)/d_v (d_v - 1), \, \mathbb{I}(x_v = x(c)))^T .
\end{equation}
When using MH, estimation proceeds by using $\mu_m$.  When using SRW,
estimation will proceed using importance sampling \citep{hest:1995,
  mcbook, robe:case:2013} with
\[
\mu_{m}^{SRW} = \frac{ \sum_{t=0}^{m-1} \left[ \frac{h(V_t)}{d_{V_t}} \right]}{\sum_{t=0}^{m-1} \frac{1}{d_{V_t}}}. 
\]
Other names for this approach include reweighted random walk or respondent driven sampling as MCMC \citep{goel2009respondent, gjoka2011practical, Avrachenkov2016, salganik2004sampling}.  To find the form of the CLT, we use a transformed version of $h$. 

Namely, let $h^*(v) = (1/d_v,\, \mathbb{I}(d_v = k)/d_v, \, 2t_v \mathbb{I}(d_v \ge 2)/d_v^2 (d_v - 1), \, \mathbb{I}(x_v = x(c))/d_v)^T$ so that if
\[
\mu^*_m= \frac{1}{m} \sum_{t=0}^{m-1} h^*(V_t),
\]

then, by the CLT, we have, as $m \to \infty$,
\[
  \sqrt{m}(\mu^*_m - \mu_{h^*}) \to \text{N}(0, \Sigma^*).
  \]
We then apply the delta method with $g(a,b,c,d)= (1/a, b/a, c/a, d/a)^T$ so that
\[
  \nabla g = \begin{pmatrix} -1/a^2 & -b/a^2 & -c/a^2 & -d/a^2\\
                              0 & 1/a & -0 & 0 \\
                              0 & 0 & 1/a & 0 \\
                              0 & 0 & 0 & 1/a
             \end{pmatrix}, 
\]
to obtain, via \eqref{eq:dm.clt}, that, as $m \to \infty$,
\[
  \sqrt{m}(g(\mu^*_m) - g(\mu_{h^*})) \to \text{N}\left(0, [\nabla g(\mu_{h^*})]^T \Sigma^* [\nabla g(\mu_{h^*})]\right) 
\]
and we can estimate the asymptotic variance with
\[
[\nabla g(\mu^*_m)]^T \Sigma_m^* [\nabla g(\mu^*_m)] .
\]

Again, the goal is to obtain estimates of these network properties and measures on the reliability of those estimates.

We now consider the algorithms and output analysis methods described
above as applied to three social networks. We begin with a simple
example to illustrate the concepts and progressively move to more
complicated, larger networks.

\section{Application to Social Networks}

To demonstrate the applicability of this work we look into classic cases in the literature: (1) a simulated network based on Ad-Health data \citep{handcock2008statnet,resnick1997protecting}, (2) a college Facebook friendship network \citep{traud2008community}, and (3) the Friendster network to showcase its use on large scale graphs. These three cases allow us to demonstrate the effectiveness of the output analysis methods.

\subsection{High School Social Network Data}\label{Toy}
The \texttt{faux.magnolia.high} social network is in the \texttt{ergm} R package \citep{handcock2008statnet,resnick1997protecting}. It is a simulation of a within-school friendship network representative of those in the southern United States. All edges are undirected and we removed 1,022 nodes out of 1,461 to ensure a well-connected graph. This resulting social network has 439 nodes (students) and 573 edges (friendships). Other nodal attributes besides structural are grade, race, and sex. The population parameters are in Tables~\ref{tab:magnoliaSummaryStats} and~\ref{tab:magnoliaSummaryStatsOther}. \par \vspace{0.5cm}

\begin{table}[ht]
\centering
\begin{tabular}{lcccccc}
  \hline
 & Min & 25\% & Median & Mean & 75\% & Max \\ 
  \hline
Degree & 1.00 & 1.00 & 2.00 & 2.61 & 4.00 & 8.00 \\ 
Triples & 0.00 & 0.00 & 1.00 & 3.15 & 6.00 & 28.00 \\ 
Triangles & 0.00 & 0.00 & 0.00 & 0.90 & 1.00 & 10.00 \\ 
Clustering Coefficient & 0.00 & 0.00 & 0.00 & 0.13 & 0.20 & 1.00\\ 
   \hline
\end{tabular} 
  \caption{Population parameters of well-connected \texttt{faux.magnolia.high} social network.}
  \label{tab:magnoliaSummaryStats}
\end{table}

\begin{table}[ht]
\centering
\begin{tabular}{lcccccc}
  \hline
Grade & Mean & SD&  &  &  & \\
    & 9.42 & 1.62 &&&\\
&&&&&&\\
Sex & Male & Female & & & & \\
  \% & 42.82 & 57.18 & & & & \\ 
&&&&&&\\   
Race & White & Black & Asian & Hisp & NatAm & Other\\
   \% & 79.73 & 12.07 & 3.19 & 3.19 & 1.37 & 0.46 \\ \hline
\end{tabular}
  \caption{Other population parameters of well-connected \texttt{faux.magnolia.high} social network.}
  \label{tab:magnoliaSummaryStatsOther}
      \end{table}
      
We ran a single chain of both the SRW and MH walks on this network with random starting nodes repeating this 1000 times independently, constructing estimates for the mean degree, mean clustering coefficient, mean grade, proportion of females, and proportion of students who identified as white. The minimum ESS for $p=5, \epsilon=0.05$, and $\alpha = 0.05$ is 10363. We also constructed the 95\% confidence region and used the corresponding volume to determine the termination time using the relative fixed-volume sequential stopping rule with multivariate batch means with the square root batch size, $\epsilon=0.05$, and $m^* = 10,000$. At this random terminating point we also noted the univariate mean estimates, multivariate effective sample size, and the number of unique nodes visited by the termination step. 

\subsubsection{Results} 
\begin{figure}[ht]
\centering
\begin{subfigure}{.45\textwidth}
  \centering
  \includegraphics[width=.9\linewidth]{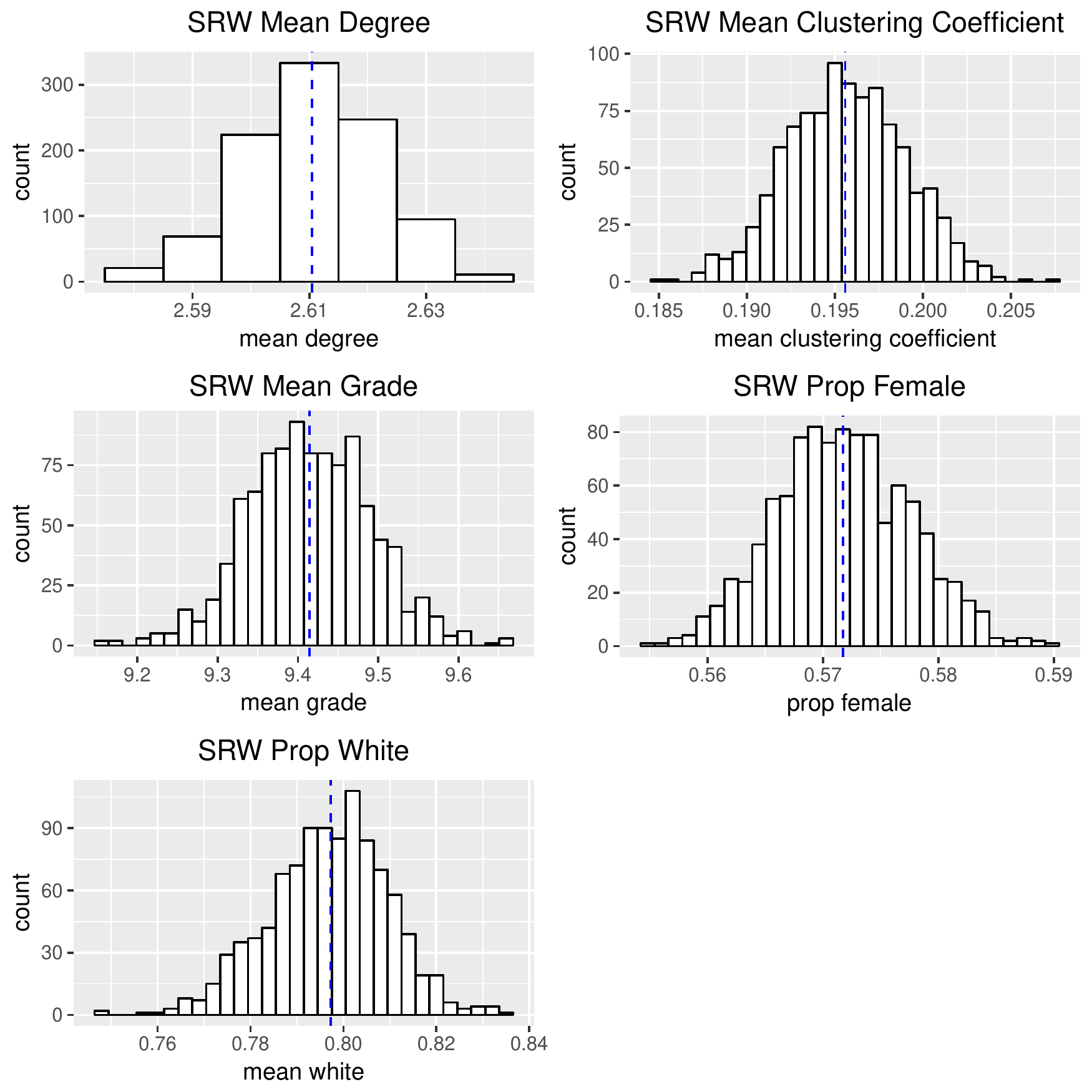}
  \caption{Mean estimates from the SRW at termination.   }
  \label{fig:SRWmeans}
\end{subfigure}%
\begin{subfigure}{.45\textwidth}
  \centering
  \includegraphics[width=.9\linewidth]{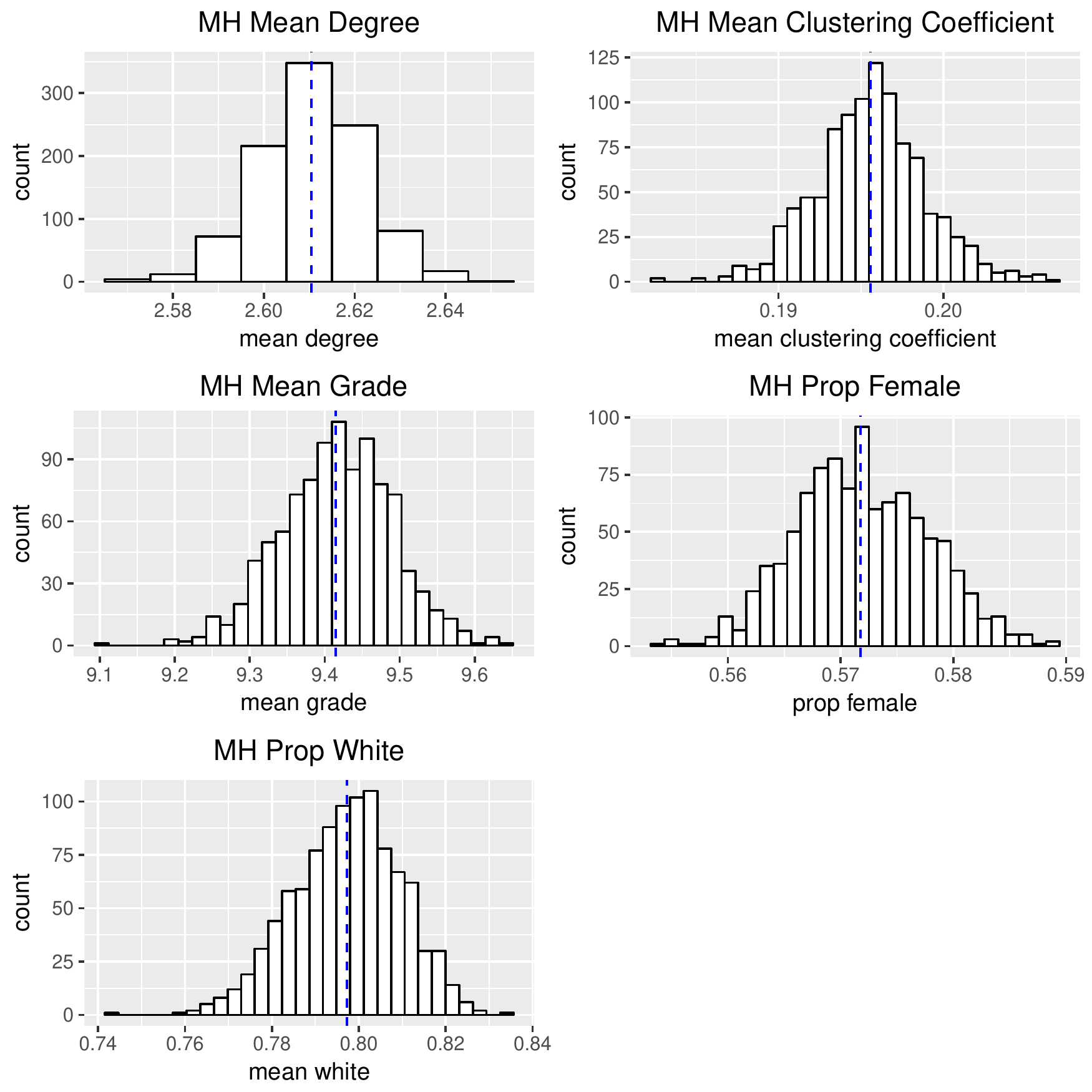}
  \caption{Mean estimates from the MH at termination.}
  \label{fig:MHmeans}
\end{subfigure}
\caption{Mean estimates from SRW and MH on well-connected \texttt{faux.magnolia.high} network. Replications = 1000. Blue dashed line indicates population quantity.}
\label{fig:MagnoliaMeanDegree}
\end{figure}

The univariate estimates with standard errors from both the SRW and MH are in Figure~\ref{fig:MagnoliaMeanDegree} and Table~\ref{tab:MagnoliaMeans}. 

\vspace{0.5cm}

\begin{table}[ht]
\centering
\begin{tabular}{rlllll}
  \hline
  Type & Degree & Clustering coeff & Grade & Prop female  & Prop white \\ 
  \hline
 Truth & 2.6105 & 0.1956 & 9.4146 & 0.5718 & 0.7973 \\ 
 SRW & 2.6106 (0.0004) & 0.1956 (0.0001) & 9.4145 (0.0026) & 0.5716 (0.0157) & 0.7969 (0.0127) \\ 
 MH & 2.6103 (0.0004) & 0.1956 (0.0001) & 9.4158 (0.0024) & 0.5719 (0.0157) & 0.7973 (0.0127) \\
   \hline
\end{tabular}
  \caption{Mean estimates from SRW and MH on the well-connected \texttt{faux.magnolia.high} network at termination time. Replications = 1000 and standard errors in parentheses.}
  \label{tab:MagnoliaMeans}
\end{table}

\begin{figure}[H]
\centering
\begin{subfigure}{.45\textwidth}
  \includegraphics[height=2in]{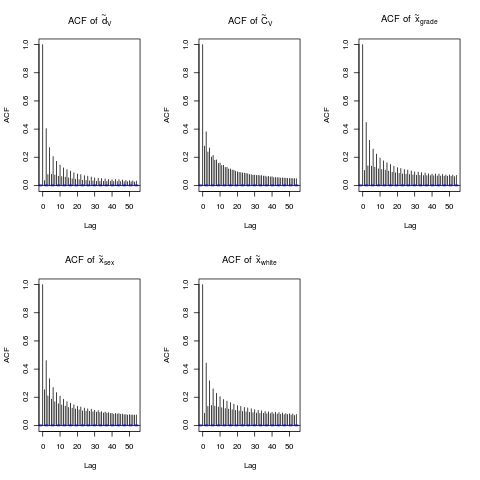}
  \caption{ACF plots from SRW.}
  \label{fig:SRWacf}
\end{subfigure}%
\begin{subfigure}{.45\textwidth}
  \includegraphics[height=2in]{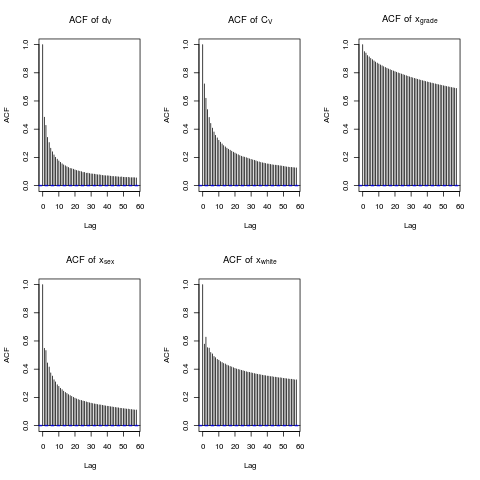}
  \caption{ACF plots from MH.}
  \label{fig:MHacf}
\end{subfigure}
\caption{ACF plots from one terminated chain of SRW and MH on \texttt{faux.magnolia.high} network.}
\label{fig:MagnoliaACFandTrace}
\end{figure}

All SRW samples terminated on average around 341,000 steps (average computer run time 425 seconds) whereas the MH samples did not achieve the stopping criterion until around 689,115 steps on average (average computer run time 352 seconds). Results are shown in Table~\ref{tab:magnoliaESS}. Since the network is relatively small, all runs of the two sampling methods captured all the nodes in the network. The mean acceptance rate of the MH samples was 0.29. Auto correlation function (ACF) plots for the five estimates from one terminated chain of the SRW and MH are shown in Figure~\ref{fig:MagnoliaACFandTrace}.

\begin{table}[H]
\centering
\begin{tabular}{rllll}
  \hline
 & Termination Step & ESS  & Unique Nodes & $T(\epsilon=0.05)$\\ 
  \hline
SRW & 341190 (452.481) & 10639.67 (3.113) & 439 (0) & 0.0497 (0) \\
MH &  689115 (698.090) & 10550.03 (2.273) & 439 (0) & 0.0498 (0) \\
   \hline
\end{tabular}
  \caption{Termination time, effective sample size, unique nodes sampled by termination for $\epsilon=0.05$, and $T(\epsilon=0.05)$ at termination step on the well-connected \texttt{faux.magnolia.high} network. Replications = 1000 and standard errors are in parentheses. }
  \label{tab:magnoliaESS}
\end{table} 

\subsection{NYU Facebook Data}
The New York University (NYU) Facebook (FB) dataset is a snapshot of anonymized Facebook data from the NYU student population in 2005 \citep{traud2008community}. Nodes are NYU FB users and edges are online friendships. The data was obtained directly from FB and is a complete set of users at NYU at the time. Other nodal attributes in this data are: gender, class year, major, high school, and residence. Some nodes had missing attribute data, so we created a new category labeled ``Not Reported'' (NR). The full NYU FB dataset contains 21,679 nodes (users) and 715,715 undirected edges (online friendships). We only considered the largest well-connected component, NYU WC FB, which has 21,623 users and 715,673 undirected edges. The population parameters of this network are in Table~\ref{tab:nyuSummaryStats}. We estimated the mean degree, mean clustering coefficient, proportion of female users, and proportion of users with major = 209. 

\begin{table}[H]
\centering
\begin{tabular}{lcccccc}
  \hline
 & Min & 25\% & Median & Mean & 75\% & Max \\ 
  \hline
Degree & 1.00 & 21.00 & 50.00 & 66.20 & 93.00 & 2315.00 \\ 
Triples & 0.00 & 210.00 & 1225.00 & 4666.47 & 4278.00 & 2678455.00 \\ 
Triangles & 0.00 & 39.00 & 197.00 & 502.24 & 598.00 & 39402.00 \\ 
Clustering Coefficient & 0.00 & 0.10 & 0.15 & 0.19 & 0.23 & 1.00\\ \hline
&&&&&&\\ \hline
Gender & Female & Male &  NR &  & \\
\% & 55.05 & 37.39 & 7.57 & & \\ 
Major & 209 & Other & NR &  & \\
\% & 6.02 & 77.82 & 16.16 & & \\ \hline
\end{tabular}
  \caption{Population parameters of well-connected NYU FB social network, NR = Not Reported. $n=21,623$, $n_e=715,673$.}
  \label{tab:nyuSummaryStats}
\end{table}

Again we ran a single chain of both the SRW and MH on this network with random starting nodes, repeating this 1000 times independently, constructing the 95\% confidence region and determining the termination time with the square root batch size, $\epsilon=0.05$ and $m^* = 10,000$. The minimum ESS for $p=4, \epsilon=0.05$, and $\alpha = 0.05$ is 9992. We constructed coverage probabilities by noting if the confidence region was below the Hotellings $T$-squared quantile. 

\subsubsection{Results} 
\begin{figure}[ht]
\centering
\begin{subfigure}{.45\textwidth}
  \centering
  \includegraphics[width=.9\linewidth]{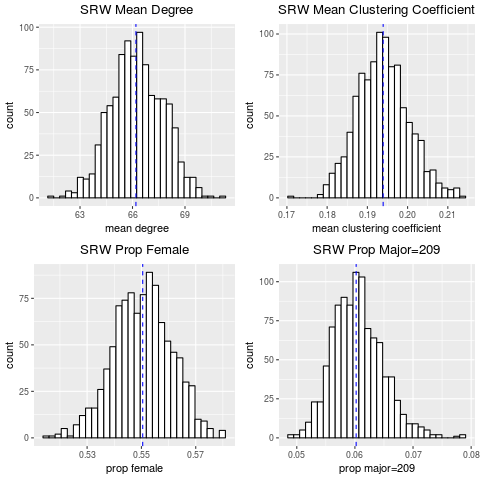}
  \caption{Mean estimates from the SRW at termination}
  \label{fig:SRWmeansNYU}
\end{subfigure}%
\begin{subfigure}{.45\textwidth}
  \centering
  \includegraphics[width=.9\linewidth]{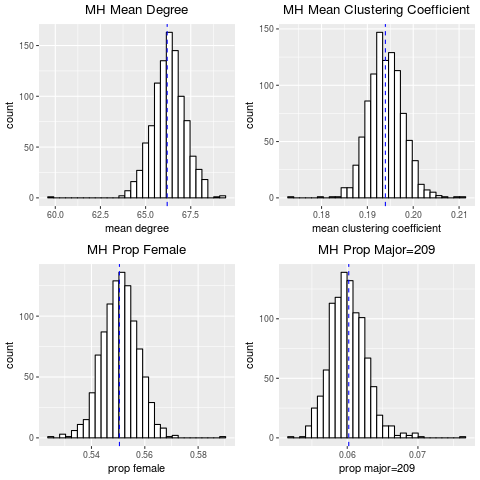}
  \caption{Mean estimates from the MH at termination}
  \label{fig:MHmeansNYU}
\end{subfigure}
\caption{Mean estimates from SRW and MH on NYU WC FB at termination. Replications = 1000. Blue dashed line indicates population quantity.}
\label{fig:NYUMeanDegree}
\end{figure}

The univariate network mean estimates are noted in Figure~\ref{fig:NYUMeanDegree} and Table~\ref{tab:NYUMeans}. The mean degree estimate from the SRW and MH on average both slightly overestimate the true mean degree. Otherwise, the estimates from both the SRW and MH algorithms are close to the population means. 

\begin{table}[ht]
\centering
\begin{tabular}{rllll}
  \hline
  Type & Degree & Clustering coeff & Prop female  & Prop Major=209 \\ 
  \hline
 Truth & 66.1955 & 0.1939 & 0.5505 & 0.0602\\
 SRW & 66.2708 (0.04714) & 0.1939 (0.0002) & 0.5504 (0.01573) & 0.0605 (0.00754) \\ 
 MH & 66.2803 (0.02853) & 0.194 (0.00012) & 0.5508 (0.0157) & 0.0601 (0.0075)  \\ 
   \hline
\end{tabular}
  \caption{Mean estimates from SRW and MH on NYC WC FB at termination time. Replications = 1000 and standard errors in parentheses.}
  \label{tab:NYUMeans}
\end{table}

\begin{table}[ht]
\centering \footnotesize
\begin{tabular}{rlllll}
  \hline
 & Termination Step & ESS  & Coverage Prob & Unique Nodes & $T(\epsilon=0.05)$\\ 
  \hline
SRW & 14676.78 (51.02) & 10558.7 (25.36) & 0.938 (0.002) & 8703.88 (17.55) & 0.048 (0.00)\\
MH &  85948.61 (416.40) & 6824.317 (11.38) & 0.91 (0.003) & 16790.81 (19.96) & 0.049 (0.00) \\
   \hline
\end{tabular}
  \caption{Termination times, effective sample size, coverage probabilities,  number of unique nodes sampled by termination time for $\epsilon = 0.05$, and $T(\epsilon=0.05)$ at termination for NYU WC FB. Replications = 1000 and standard errors in parentheses.}
  \label{tab:nyuESS}
\end{table}

\begin{figure}[ht]
\centering
\begin{subfigure}{.5\textwidth}
  \centering
  \includegraphics[width=.95\linewidth]{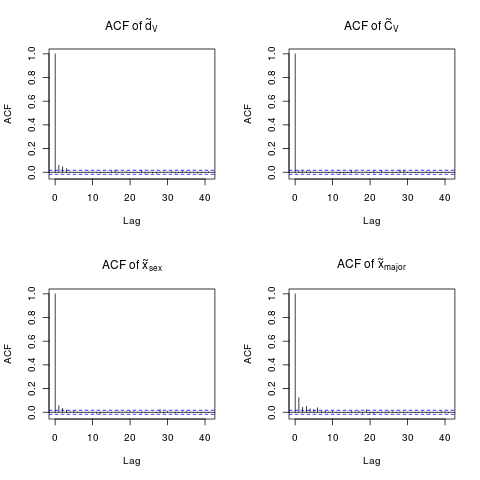}
  \caption{ACF plots from SRW.}
  \label{fig:SRWacfNYU}
\end{subfigure}%
\begin{subfigure}{.5\textwidth}
  \centering
  \includegraphics[width=.95\linewidth]{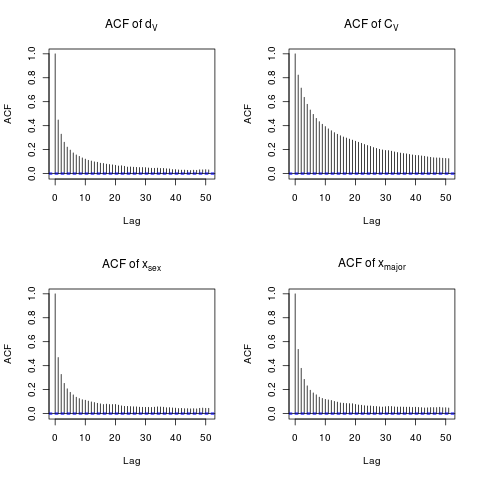}
  \caption{ACF plots from MH.}
  \label{fig:MHacfNYU}
\end{subfigure}
\caption{ACF plots from one chain of SRW and MH on NYU WC FB network.}
\label{fig:NYUacfAndTrace}
\end{figure}

All SRW samples terminated on average around 14,700 steps (average computer run time 8.1 seconds) whereas among the MH samples terminated on average by 86,000 steps (average computer run time 30.9 seconds), see Table~\ref{tab:nyuESS}. The mean acceptance rate of the MH walks was 0.5621. ACF plots for one chain of both the SRW and MH are shown in Figure~\ref{fig:NYUacfAndTrace}.

\newpage
\subsection{Friendster Data}
The Friendster dataset is hosted on the Stanford Large Network Dataset (SNAP) web site \citep{leskovec2016snap}. Friendster was an online social gaming and social networking site, where members had user profiles and could link to one another. Friendster also allowed users to form groups which other members could join. The SNAP-hosted Friendster dataset is the largest well-connected component of the induced subgraph of nodes that belonged to at least one group or were connected to other nodes that belonged to at least one group. This social network has 65,608,366 nodes (users) and 1,806,067,135 undirected edges (friendships). There are no other nodal attributes in this data. We estimated the mean degree and mean clustering coefficient. 

  \subsubsection{Implementation}
  We ran 100 chains of length 100,000 from random starting nodes. To find these random starting nodes we generated random numbers and searched if it existed in the network. If it existed, the sample began at this node, if not we generated another random number until it was accepted. During the sampling procedure we collected the visited node's id, neighborhood, and calculated its degree. Running all 100 independent chains on five cores, took around 80 minutes for the SRW samples and 116 minutes for the MH samples. After completing the walks, we queried the file again to count the number of triangles for each visited node. Counting triangles is a computationally expensive step, so we only computed triangles on the chains up to length 10,000. Therefore, the multivariate results we present are on shorter chains of length 10,000, but we also present full 100,000 results on the univariate estimate of mean degree. 

\subsubsection{Shorter chain results}
\begin{figure}[h]
  \centering
  \includegraphics[scale=0.6]{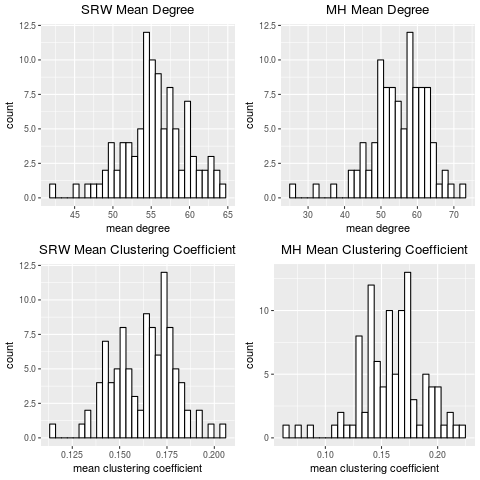}
  \caption{Mean estimates from SRW and MH walks on the Friendster network for 10,000 length chains. Replications = 100.}
  \label{fig:friendstermean1e4}
\end{figure}

Results are in Figure~\ref{fig:friendstermean1e4} and Tables~\ref{tab:FriendsterMeans} and~\ref{tab:FriendsterESS1e4}. The mean degree estimate from both the SRW and MH is around 55 with more variability in the MH samples and the mean clustering coefficient for both algorithms is around 0.16. 

\begin{table}[h]
\centering
\begin{tabular}{rll}
  \hline
  Type & Degree & Clustering coeff  \\ 
  \hline
 SRW & 55.51 (0.414) & 0.163 (0.002) \\ 
 MH & 54.97 (0.765) & 0.159 (0.009) \\ 
   \hline
\end{tabular}
  \caption{Mean estimates from the SRW and MH on Friendster network with chain length 10,000. Replications = 100 and standard errors in parentheses.}
  \label{tab:FriendsterMeans}
\end{table}

\begin{figure}[h]
\centering
\begin{subfigure}{.5\textwidth}
  \centering
  \includegraphics[width=.9\linewidth]{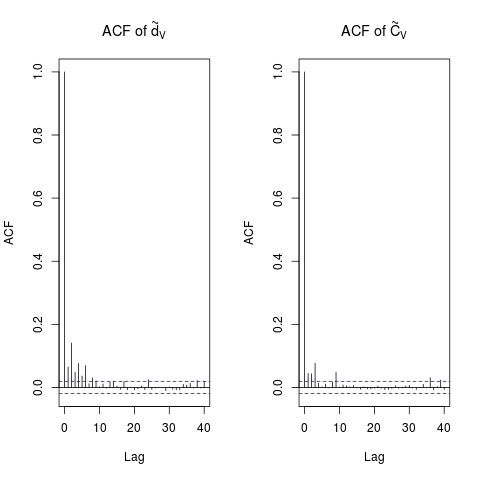}
  \caption{ACF plots from SRW.}
  \label{fig:SRWacfFriendster1e4}
\end{subfigure}%
\begin{subfigure}{.5\textwidth}
  \centering
  \includegraphics[width=.9\linewidth]{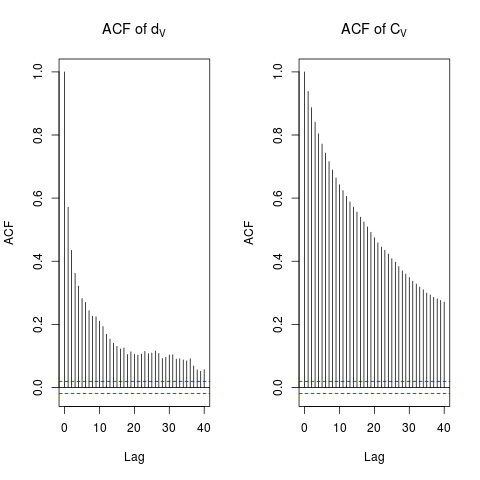}
  \caption{ACF plots from MH.}
  \label{fig:MHacfFriendster1e4}
\end{subfigure}
\caption{ACF plots from one 1e4 chain of SRW and MH on Friendster network.}
\label{fig:Friendster1e4acfAndTrace}
\end{figure}

The striking difference between the SRW and MH is in the effective sample size and number of unique nodes captured. The MH walks on average collect only around 25\% of the unique nodes that the SRW does. And in the multivariate ESS, the MH on average is less than 20\% of the SRW. The mean acceptance rate in the MH walks was 0.2904. The minimum ESS for $p=2, \epsilon=0.05$, and $\alpha=0.05$ is 7530, where none of the simulations achieved the minimum ESS for reliable estimation by 10,000 steps. This implies more samples are needed. ACF plots for one chain are shown in Figure~\ref{fig:Friendster1e4acfAndTrace}. 
\begin{table}[H]
\centering
\begin{tabular}{rlll}
  \hline
  & $T(\epsilon=0.05)$ & ESS  & Unique Nodes\\ 
  \hline
SRW & 0.058 (0.0004) & 3865.95 (212.399)  & 9797 (2.096)\\
MH & 0.0985 (0.0002) & 462.918 (6.467)  &  2437 (27.023)\\
   \hline
\end{tabular}
  \caption{Multivariate: $T_{SD}(\epsilon=0.05)$, effective sample size, and number of unique nodes sampled by 10,000 steps in Friendster network. Replications = 100 and standard errors in parentheses.}
  \label{tab:FriendsterESS1e4}
\end{table}

\subsubsection{Full chain results}
\begin{figure}[h]
  \centering
  \includegraphics[scale=0.4]{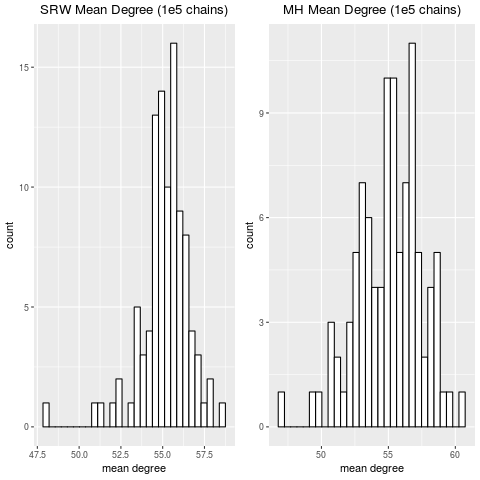}
  \caption{Mean estimates from SRW and MH walks on the Friendster network for 100,000 length chains. Replications = 100.}
  \label{fig:friendstermeandegree1e5t}
\end{figure}

\begin{figure}[H]
\centering
\begin{subfigure}{.5\textwidth}
  \centering
  \includegraphics[width=.95\linewidth]{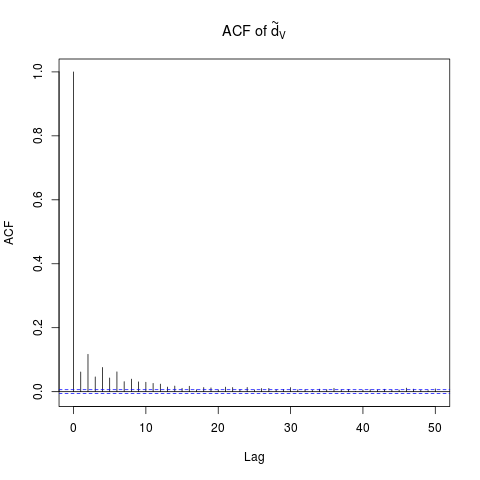}
  \caption{ACF plots from SRW.}
  \label{fig:SRWacfFriendster1e5}
\end{subfigure}%
\begin{subfigure}{.5\textwidth}
  \centering
  \includegraphics[width=.95\linewidth]{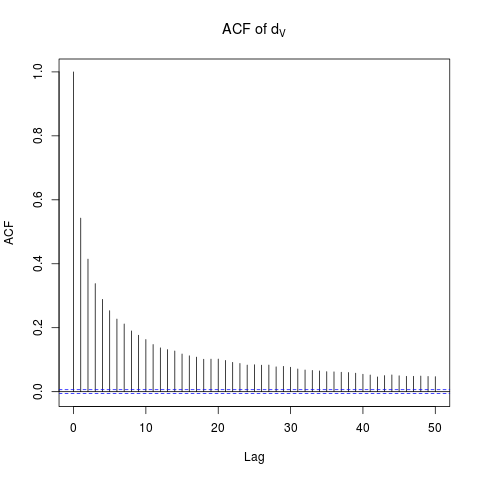}
  \caption{ACF plots from MH.}
  \label{fig:MHacfFriendster1e5}
\end{subfigure}
\caption{ACF plots from one 1e5 chain of SRW and MH on Friendster network.}
\label{fig:Friendster1e5acfAndTrace}
\end{figure}

If we consider estimating the mean degree of the 100,000 length chains, we see the mean degree estimates from the SRW and MH walks are again similar. Likewise, the ESS and number of unique nodes are on starkly different scales (Figure~\ref{fig:friendstermeandegree1e5t} and Table~\ref{tab:FriendsterESS1e5}). We use the result from Proposition~\ref{prop:dESS}, with $p=1$, $g(x) = 1/x$ and the square root batch means estimation to calculate the univariate ESS. The mean acceptance rate of of the MH walks was 0.2905.  ACF plots for one chain are shown in Figure~\ref{fig:Friendster1e5acfAndTrace}. 

\begin{table}[H]
\centering
\begin{tabular}{rlll}
  \hline
 & Degree & ESS  & Unique Nodes\\ 
  \hline
SRW & 55.15 (0.149)  & 36229 (1408.53) & 97474 (14.124)\\
MH &  55.07 (0.245) & 6002 (53.507)  & 24477 (91.33) \\
   \hline
\end{tabular}
  \caption{Univariate: mean degree, effective sample size, and number of unique nodes sample by 100,000 steps for $\epsilon=0.05$ for Friendster network. Replications = 100 and standard errors in parenthesis.}
  \label{tab:FriendsterESS1e5}
\end{table}

\subsection{Summary of results}
Consistently across all three networks, the SRW was more efficient than the MH, either with respect to the termination time to achieve the stopping criterion or with respect to the effective sample size. Our results confirm what other authors have found in univariate settings \citep{gjoka2011practical,Avrachenkov2016}. In addition, as clearly indicated in the histograms, repeated runs of the algorithms obtained slightly different estimates. However, when the minimum effective sample size was reached, the variation in these estimates was small. This further emphasizes that prior to running the algorithms on any of these networks, a researcher can determine the simulation effort required via the minimum ESS. Once that minimum ESS has been reached, researchers will have an approximately $100(1-\alpha)$\% confidence with precision $\epsilon$ for the $p$ many estimates (as shown in Table~\ref{tab:MinESS}).  

\begin{table}[h]
  \centering
  \begin{tabular}{cccc}\hline
  $p$ & Conf level & $\epsilon$ & Minimum ESS\\ \hline
  5 & 95\% & 0.05 & 10363\\
  4 & 95\% & 0.05 & 9992\\
  2 & 95\% & 0.05 & 7530\\ \hline
  \end{tabular}
  \caption{Minimum ESS required for $p$ estimated features at a $100(1-\alpha)$\% confidence level and threshold level $\epsilon$.}
  \label{tab:MinESS}
\end{table}


\section{Discussion}
The use of MCMC methods on networks without sampling frames to estimate multiple features is common. However, the error associated with the estimation in the multivariate setting has not been studied closely. We contribute to the literature by further developing multivariate MCMC output analysis methods in the context of network sampling that directly addresses the reliability of the multivariate estimation. 

We support existing findings that the MH is less efficient than the SRW in univariate estimation and extend the results to a multivariate setting. We have also extended the MCMC output analysis framework more generally so that it can be applied to other MCMC algorithms. If a researcher plans to use an MCMC method to collect a sample, they can now find the minimum number of effective samples they should collect before they terminate the sampling procedure. Moreover, they have the tools to assess the reliability of the inference they make from that sample. By using such tools, researchers can have greater confidence in the consistency and reproducibility of their results. This reduces the chance of outlier results or non-reproducible estimates due to insufficient Monte Carlo sample sizes. 

There are multiple extensions of this work that could benefit from further research. First, it would be interesting to extend this research to handle edge sampling algorithms to estimate network edge properties. In addition, we focused on binary networks, so generalizing the framework to work on weighted networks that convey relationship strength or weakness would be useful. Another extension is to develop these methods to work on directed networks. The most practically beneficial extension, though, may be to use these reliable estimation tools, such as minimum effective sample size, in the context of RDS. However, the assumptions required for the output analysis tools are not met in RDS, therefore further work is required to apply the methods we propose.

\bibliographystyle{apalike}
\bibliography{mybibfile}

\begin{thebibliography}{}

\bibitem[Ahmed et~al., 2014]{ahmed2014network}
Ahmed, N.~K., Neville, J., and Kompella, R. (2014).
\newblock Network sampling: From {S}tatic to {S}treaming {G}raphs.
\newblock {\em ACM Transactions on Knowledge Discovery from Data (TKDD)},
  8(2):7.

\bibitem[Aldous et~al., 1997]{aldous1997mixing}
Aldous, D., Lov{\'a}sz, L., and Winkler, P. (1997).
\newblock Mixing times for uniformly ergodic {M}arkov chains.
\newblock {\em Stochastic Processes and their Applications}, 71(2):165--185.

\bibitem[Andrews, 1991]{andr:1991}
Andrews, D. (1991).
\newblock Heteroskedasticity and autocorrelation consistent covariant matrix
  estimation.
\newblock {\em Econometrica}, 59(3):817--858.

\bibitem[Avrachenkov et~al., 2016]{Avrachenkov2016}
Avrachenkov, K., Borkar, V.~S., Kadavankandy, A., and Sreedharan, J.~K. (2016).
\newblock Comparison of {R}andom {W}alk {B}ased {T}echniques for {E}stimating
  {N}etwork {A}verages.
\newblock In Nguyen, H.~T. and Snasel, V., editors, {\em Computational Social
  Networks: 5th International Conference, CSoNet 2016, Ho Chi Minh City,
  Vietnam, August 2-4, 2016, Proceedings}, pages 27--38, Cham. Springer
  International Publishing.

\bibitem[Avrachenkov et~al., 2018]{avrachenkov2018revisiting}
Avrachenkov, K., Borkar, V.~S., Kadavankandy, A., and Sreedharan, J.~K. (2018).
\newblock Revisiting random walk based sampling in networks: evasion of burn-in
  period and frequent regenerations.
\newblock {\em Computational Social Networks}, 5(1):4.

\bibitem[Blagus et~al., 2017]{blagus2017empirical}
Blagus, N., {\v{S}}ubelj, L., and Bajec, M. (2017).
\newblock Empirical comparison of network sampling: {H}ow to choose the most
  appropriate method?
\newblock {\em Physica A: Statistical Mechanics and its Applications},
  477:136--148.

\bibitem[Bremaud, 2010]{bremaud_2010}
Bremaud, P. (2010).
\newblock {\em Markov {C}hains: {G}ibbs {F}ields, {M}onte {C}arlo {S}imulation,
  and {Q}ueues}.
\newblock Springer Science and Business Media.

\bibitem[Chen and Seila, 1987]{chen1987multivariate}
Chen, D.-F.~R. and Seila, A.~F. (1987).
\newblock Multivariate inference in stationary simulation using batch means.
\newblock In {\em Proceedings of the 19th conference on Winter simulation},
  pages 302--304. ACM.

\bibitem[Chiericetti et~al., 2016]{chiericetti2016sampling}
Chiericetti, F., Dasgupta, A., Kumar, R., Lattanzi, S., and Sarl{\'o}s, T.
  (2016).
\newblock On sampling nodes in a network.
\newblock In {\em Proceedings of the 25th International Conference on World
  Wide Web}, pages 471--481. International World Wide Web Conferences Steering
  Committee.

\bibitem[Cowles and Carlin, 1996]{cowl:carl:1996}
Cowles, M.~K. and Carlin, B.~P. (1996).
\newblock Markov chain {M}onte {C}arlo convergence diagnostics: {A} comparative
  review.
\newblock {\em Journal of the American Statistical Association}, 91:883--904.

\bibitem[Dai and Jones, 2017]{dai2017multivariate}
Dai, N. and Jones, G.~L. (2017).
\newblock Multivariate initial sequence estimators in {M}arkov chain {M}onte
  {C}arlo.
\newblock {\em Journal of {M}ultivariate {A}nalysis}, 159:184--199.

\bibitem[Flegal et~al., 2015]{flegal2015mcmcse}
Flegal, J., Hughes, J., and Vats, D. (2015).
\newblock mcmcse: {M}onte {C}arlo standard errors for {MCMC}.
\newblock {\em Riverside, CA and Minneapolis, MN. R package version 1.3.2}.

\bibitem[Flegal et~al., 2008]{fleg:etal:2008}
Flegal, J.~M., Haran, M., and Jones, G.~L. (2008).
\newblock {M}arkov chain {M}onte {C}arlo: {C}an we trust the third significant
  figure?
\newblock {\em Statistical Science}, 23:250--260.

\bibitem[Gelman and Rubin, 1992]{gelm:rubi:1992}
Gelman, A. and Rubin, D.~B. (1992).
\newblock Inference from iterative simulation using multiple sequences (with
  discussion).
\newblock {\em Statistical Science}, 7:457--472.

\bibitem[Geweke, 1992]{gewe:1992}
Geweke, J. (1992).
\newblock Evaluating the accuracy of sampling-based approaches to the
  calculation of posterior moments (with discussion).
\newblock In Bernardo, J.~M., Berger, J.~O., Dawid, A.~P., and Smith, A. F.~M.,
  editors, {\em Bayesian Statistics 4. Proceedings of the Fourth Valencia
  International Meeting}, pages 169--188. Clarendon Press.

\bibitem[Gile and Handcock, 2010]{gile2010respondent}
Gile, K.~J. and Handcock, M.~S. (2010).
\newblock Respondent-driven {S}ampling: an {A}ssessment of {C}urrent
  {M}ethodology.
\newblock {\em Sociological Methodology}, 40(1):285--327.

\bibitem[Gjoka et~al., 2011a]{gjoka2011multigraph}
Gjoka, M., Butts, C.~T., Kurant, M., and Markopoulou, A. (2011a).
\newblock Multigraph sampling of online social networks.
\newblock {\em IEEE Journal on Selected Areas in Communications},
  29(9):1893--1905.

\bibitem[Gjoka et~al., 2010]{gjoka2010walking}
Gjoka, M., Kurant, M., Butts, C.~T., and Markopoulou, A. (2010).
\newblock Walking in {F}acebook: A {C}ase {S}tudy of {U}nbiased {S}ampling of
  {OSN}s.
\newblock In {\em Infocom, 2010 Proceedings IEEE}, pages 1--9. IEEE.

\bibitem[Gjoka et~al., 2011b]{gjoka2011practical}
Gjoka, M., Kurant, M., Butts, C.~T., and Markopoulou, A. (2011b).
\newblock Practical {R}ecommendations on {C}rawling {O}nline {S}ocial
  {N}etworks.
\newblock {\em IEEE Journal on Selected Areas in Communications},
  29(9):1872--1892.

\bibitem[Glynn and Whitt, 1992]{glynnwhitt:1992}
Glynn, P.~W. and Whitt, W. (1992).
\newblock The asymptotic efficiency of simulation estimators.
\newblock {\em Operations research}, 40(3):505--520.

\bibitem[Goel and Salganik, 2009]{goel2009respondent}
Goel, S. and Salganik, M.~J. (2009).
\newblock Respondent-driven sampling as {M}arkov chain {M}onte {C}arlo.
\newblock {\em Statistics in Medicine}, 28(17):2202--2229.

\bibitem[Handcock and Gile, 2010]{handcock2010modeling}
Handcock, M.~S. and Gile, K.~J. (2010).
\newblock Modeling social networks from sampled data.
\newblock {\em The Annals of Applied Statistics}, 4(1):5.

\bibitem[Handcock et~al., 2008]{handcock2008statnet}
Handcock, M.~S., Hunter, D.~R., Butts, C.~T., Goodreau, S.~M., and Morris, M.
  (2008).
\newblock statnet: {S}oftware {T}ools for the {R}epresentation,
  {V}isualization, {A}nalysis and {S}imulation of {N}etwork {D}ata.
\newblock {\em Journal of Statistical Software}, 24(1):1548.

\bibitem[Heckathorn, 1997]{heckathorn1997respondent}
Heckathorn, D.~D. (1997).
\newblock Respondent-driven sampling: a new approach to the study of hidden
  populations.
\newblock {\em Social problems}, 44(2):174--199.

\bibitem[Heidelberger and Welch, 1983]{heid:welc:1983}
Heidelberger, P. and Welch, P.~D. (1983).
\newblock Simulation run length control in the presence of an initial
  transient.
\newblock {\em Operations Research}, 31:1109--1144.

\bibitem[Hesterberg, 1995]{hest:1995}
Hesterberg, T. (1995).
\newblock Weighted average importance sampling and defensive mixture
  distributions.
\newblock {\em Technometrics}, 37:185--194.

\bibitem[Hobert et~al., 2002]{hobe:etal:2002}
Hobert, J.~P., Jones, G.~L., Presnell, B., and Rosenthal, J.~S. (2002).
\newblock On the applicability of regenerative simulation in {M}arkov chain
  {M}onte {C}arlo.
\newblock {\em Biometrika}, 89:731--743.

\bibitem[Jones, 2004]{jones2004markov}
Jones, G.~L. (2004).
\newblock On the {M}arkov chain central limit theorem.
\newblock {\em Probability Surveys}, 1:299--320.

\bibitem[Jones et~al., 2006a]{jone:etal:2006}
Jones, G.~L., Haran, M., Caffo, B.~S., and Neath, R. (2006a).
\newblock Fixed-width output analysis for {M}arkov chain {M}onte {C}arlo.
\newblock {\em Journal of the American Statistical Association},
  101:1537--1547.

\bibitem[Jones et~al., 2006b]{jones2006fixed}
Jones, G.~L., Haran, M., Caffo, B.~S., and Neath, R. (2006b).
\newblock Fixed-{W}idth {O}utput {A}nalysis for {M}arkov {C}hain {M}onte
  {C}arlo.
\newblock {\em Journal of the American Statistical Association},
  101(476):1537--1547.

\bibitem[Joyce, 2011]{joyce2011kullback}
Joyce, J.~M. (2011).
\newblock Kullback-{L}eibler {D}ivergence.
\newblock In {\em International Encyclopedia of Statistical Science}, pages
  720--722. Springer.

\bibitem[Kolaczyk, 2009]{Kolaczyk2009}
Kolaczyk, E.~D. (2009).
\newblock {\em Statistical {A}nalysis of {N}etwork {D}ata: {M}ethods and
  {M}odels}.
\newblock Springer New York.

\bibitem[Kosorok, 2000]{kosorok2000monte}
Kosorok, M.~R. (2000).
\newblock Monte {C}arlo error estimation for multivariate {M}arkov chains.
\newblock {\em Statistics \& probability letters}, 46(1):85--93.

\bibitem[Kurant et~al., 2012]{kurant2012coarse}
Kurant, M., Gjoka, M., Wang, Y., Almquist, Z.~W., Butts, C.~T., and
  Markopoulou, A. (2012).
\newblock Coarse-grained topology estimation via graph sampling.
\newblock In {\em Proceedings of the 2012 ACM workshop on Workshop on online
  social networks}, pages 25--30. ACM.

\bibitem[Lee et~al., 2012]{lee2012beyond}
Lee, C.-H., Xu, X., and Eun, D.~Y. (2012).
\newblock Beyond {R}andom {W}alk and {M}etropolis-{H}astings {S}amplers: {W}hy
  {Y}ou {S}hould {N}ot {B}acktrack for {U}nbiased {G}raph {S}ampling.
\newblock In {\em ACM SIGMETRICS Performance {E}valuation {R}eview}, volume~40,
  pages 319--330. ACM.

\bibitem[Lee et~al., 2006]{lee2006statistical}
Lee, S.~H., Kim, P.-J., and Jeong, H. (2006).
\newblock Statistical properties of sampled networks.
\newblock {\em Physical Review E}, 73(016102):1--7.

\bibitem[Leskovec and Faloutsos, 2006]{leskovec2006sampling}
Leskovec, J. and Faloutsos, C. (2006).
\newblock Sampling from {L}arge {G}raphs.
\newblock In {\em Proceedings of the 12th ACM SIGKDD {I}nternational
  {C}onference on Knowledge Discovery and Data Mining}, pages 631--636. ACM.

\bibitem[Leskovec and Sosi{\v{c}}, 2016]{leskovec2016snap}
Leskovec, J. and Sosi{\v{c}}, R. (2016).
\newblock {SNAP}: A {G}eneral-{P}urpose {N}etwork {A}nalysis and
  {G}raph-{M}ining {L}ibrary.
\newblock {\em ACM Transactions on Intelligent Systems and Technology (TIST)},
  8(1):1.

\bibitem[Levin et~al., 2009]{levin2009markov}
Levin, D.~A., Peres, Y., and Wilmer, E.~L. (2009).
\newblock {\em Markov {C}hains and {M}ixing {T}imes}.
\newblock American Mathematical Soc.

\bibitem[Li et~al., 2015]{li2015random}
Li, R.-H., Yu, J.~X., Qin, L., Mao, R., and Jin, T. (2015).
\newblock On {R}andom {W}alk {B}ased {G}raph {S}ampling.
\newblock In {\em Data Engineering (ICDE), 2015 IEEE 31st International
  Conference on}, pages 927--938. IEEE.

\bibitem[Liu and Flegal, 2018a]{liu:fleg:2018spec}
Liu, Y. and Flegal, J. (2018a).
\newblock Optimal mean squared error bandwidth for spectral variance estimators
  in mcmc simulations.
\newblock {\em ArXiv e-prints}.

\bibitem[Liu and Flegal, 2018b]{liu:fleg:2018}
Liu, Y. and Flegal, J.~M. (2018b).
\newblock Weighted batch means estimators in {M}arkov chain {M}onte {C}arlo.
\newblock {\em Electron. J. Statist.}, 12(2):3397--3442.

\bibitem[Lu and Li, 2012]{lu2012sampling}
Lu, J. and Li, D. (2012).
\newblock Sampling online social networks by random walk.
\newblock In {\em Proceedings of the First ACM International Workshop on Hot
  Topics on Interdisciplinary Social Networks Research}, pages 33--40. ACM.

\bibitem[Marsden, 2011]{marsden2011survey}
Marsden, P.~V. (2011).
\newblock Survey methods for network data.
\newblock {\em The SAGE handbook of social network analysis}, 25:370--388.

\bibitem[Mohaisen et~al., 2010]{mohaisen:2010}
Mohaisen, A., Yun, A., and Kim, Y. (2010).
\newblock Measuring the mixing time of social graphs.
\newblock In {\em Proceedings of the 10th ACM SIGCOMM conference on Internet
  measurement}, pages 383--389. ACM.

\bibitem[Owen, 2013]{mcbook}
Owen, A.~B. (2013).
\newblock {\em Monte {C}arlo theory, methods and examples}.

\bibitem[Resnick et~al., 1997]{resnick1997protecting}
Resnick, M.~D., Bearman, P.~S., Blum, R.~W., Bauman, K.~E., Harris, K.~M.,
  Jones, J., Tabor, J., Beuhring, T., Sieving, R.~E., Shew, M., et~al. (1997).
\newblock Protecting adolescents from harm: {F}indings from the {N}ational
  {L}ongitudinal {S}tudy on {A}dolescent {H}ealth.
\newblock {\em Journal of the American Medical Association}, 278(10):823--832.

\bibitem[Ribeiro and Towsley, 2010]{ribeiro:2010}
Ribeiro, B. and Towsley, D. (2010).
\newblock Estimating and sampling graphs with multidimensional random walks.
\newblock In {\em Proceedings of the 10th ACM SIGCOMM conference on Internet
  measurement}, pages 390--403. ACM.

\bibitem[Robert and Casella, 2013]{robe:case:2013}
Robert, C.~P. and Casella, G. (2013).
\newblock {\em Monte Carlo Statistical Methods}.
\newblock Springer, New York.

\bibitem[Salamanos et~al., 2017]{salamanos2017deterministic}
Salamanos, N., Voudigari, E., and Yannakoudakis, E.~J. (2017).
\newblock Deterministic graph exploration for efficient graph sampling.
\newblock {\em Social Network Analysis and Mining}, 7(1):24.

\bibitem[Salganik and Heckathorn, 2004]{salganik2004sampling}
Salganik, M.~J. and Heckathorn, D.~D. (2004).
\newblock Sampling and estimation in hidden populations using
  {R}espondent-{D}riven {S}ampling.
\newblock {\em Sociological {M}ethodology}, 34(1):193--240.

\bibitem[Scott, 2017]{scott2017social}
Scott, J. (2017).
\newblock {\em Social {N}etwork {A}nalysis}.
\newblock Sage.

\bibitem[Seila, 1982]{seil:1982}
Seila, A.~F. (1982).
\newblock Multivariate estimation in regenerative simulation.
\newblock {\em Operations Research Letters}, 1:153--156.

\bibitem[Sen and Singer, 1993]{sen:sing:1993}
Sen, P.~K. and Singer, J. d.~M. (1993).
\newblock {\em Large Sample Methods in Statistics: an Introduction with
  Applications}.
\newblock Chapman \& Hall Ltd, London.

\bibitem[Thompson, 2017]{thompson:2017}
Thompson, S.~K. (2017).
\newblock Adaptive and {N}etwork {S}ampling for {I}nference and {I}nterventions
  in {C}hanging {P}opulations.
\newblock {\em Journal of {S}urvey {S}tatistics and {M}ethodology}, 5(1):1--21.

\bibitem[Traud et~al., 2008]{traud2008community}
Traud, A.~L., Kelsic, E.~D., Mucha, P.~J., and Porter, M.~A. (2008).
\newblock Community {S}tructure in {O}nline {C}ollegiate {S}ocial {N}etworks.
\newblock {\em arXiv preprint arXiv:0809.0960}.

\bibitem[Vats and Flegal, 2018]{vats:fleg:2018}
Vats, D. and Flegal, J.~M. (2018).
\newblock Lugsail lag windows and their application to {MCMC}.
\newblock {\em arXiv:1809.04541}.

\bibitem[Vats et~al., 2019]{vats2015multivariate}
Vats, D., Flegal, J.~M., and Jones, G.~L. (2019).
\newblock Multivariate output analysis for {M}arkov chain {M}onte {C}arlo.
\newblock {\em Biometrika}, 106(2):321--337.

\bibitem[Vats et~al., 2018]{vats2018strong}
Vats, D., Flegal, J.~M., Jones, G.~L., et~al. (2018).
\newblock Strong consistency of multivariate spectral variance estimators in
  {M}arkov chain{ M}onte {C}arlo.
\newblock {\em Bernoulli}, 24(3):1860--1909.

\bibitem[Vats and Knudson, 2018]{vats:2018revisiting}
Vats, D. and Knudson, C. (2018).
\newblock Revisiting the {G}elman-{R}ubin {D}iagnostic.
\newblock {\em arXiv preprint arXiv:1812.09384}.

\bibitem[Wang et~al., 2011]{wang2011understanding}
Wang, T., Chen, Y., Zhang, Z., Xu, T., Jin, L., Hui, P., Deng, B., and Li, X.
  (2011).
\newblock Understanding {G}raph {S}ampling {A}lgorithms for {S}ocial {N}etwork
  {A}nalysis.
\newblock In {\em Distributed Computing Systems Workshops (ICDCSW), 2011 31st
  International Conference}, pages 123--128. IEEE.

\bibitem[Wasserman and Faust, 1994]{wasserman1994social}
Wasserman, S. and Faust, K. (1994).
\newblock {\em Social {N}etwork {A}nalysis: {M}ethods and {A}pplications},
  volume~8.
\newblock Cambridge {U}niversity {P}ress.

\bibitem[Wilks, 1932]{wilk:1932}
Wilks, S.~S. (1932).
\newblock Certain generalizations in the analysis of variance.
\newblock {\em Biometrika}, 24(3-4):471--494.

\bibitem[Zhou et~al., 2016]{zhou2016faster}
Zhou, Z., Zhang, N., Gong, Z., and Das, G. (2016).
\newblock Faster {R}andom {W}alks by {R}ewiring {O}nline {S}ocial {N}etworks
  {O}n-the-{F}ly.
\newblock {\em ACM Transactions on Database Systems (TODS)}, 40(4):26.

\end{thebibliography}
\end{document}